\documentclass{emulateapj}
\newcommand{\etal}{et~al.}
\newcommand{\Lsun}{L$_{\odot}$}

\newcommand{\cgsflux}{erg~s$^{-1}$~cm$^{-2}$}
\newcommand{\cmsq}{\hbox{cm$^{-2}$}}

\newcommand{\ArII}{\hbox{{\rm Ar}\kern 0.1em{\sc ii}}}
\newcommand{\ArIII}{\hbox{{\rm Ar}\kern 0.1em{\sc iii}}}
\newcommand{\CIV}{\hbox{{\rm C}\kern 0.1em{\sc iv}}}
\newcommand{\HI}{\hbox{{\rm H}\kern 0.1em{\sc i}}}
\newcommand{\HII}{\hbox{{\rm H}\kern 0.1em{\sc ii}}}
\newcommand{\HeI}{\hbox{{\rm He}\kern 0.1em{\sc i}}}
\newcommand{\HeII}{\hbox{{\rm He}\kern 0.1em{\sc ii}}}
\newcommand{\NII}{\hbox{{\rm N}\kern 0.1em{\sc ii}}}
\newcommand{\OI}{\hbox{{\rm O}\kern 0.1em{\sc i}}}
\newcommand{\OII}{\hbox{{\rm O}\kern 0.1em{\sc ii}}}
\newcommand{\OIII}{\hbox{{\rm O}\kern 0.1em{\sc iii}}}
\newcommand{\OIIlong}{{\rm O}\kern 0.1em{\sc ii}~$\lambda 3727$} 
\newcommand{\FeII}{\hbox{{\rm Fe}\kern 0.1em{\sc ii}}}
\newcommand{\NeII}{\hbox{{\rm Ne}\kern 0.1em{\sc ii}}}
\newcommand{\NeIII}{\hbox{{\rm Ne}\kern 0.1em{\sc iii}}}
\newcommand{\NeV}{\hbox{{\rm Ne}\kern 0.1em{\sc v}}}
\newcommand{\SII}{\hbox{{\rm S}\kern 0.1em{\sc ii}}}
\newcommand{\SIII}{\hbox{{\rm S}\kern 0.1em{\sc iii}}}
\newcommand{\SIV}{\hbox{{\rm S}\kern 0.1em{\sc iv}}}
\newcommand{\SiIV}{\hbox{{\rm Si}\kern 0.1em{\sc iv}}}
\newcommand{\MgII}{\hbox{{\rm Mg}\kern 0.1em{\sc ii}}}
\newcommand{\Halpha}{\hbox{{\rm H}\kern 0.1em$\alpha$}}
\newcommand{\Hbeta}{\hbox{{\rm H}\kern 0.1em$\beta$}}
\newcommand{\Heopta}{\hbox{{\rm He}\kern 0.1em{\sc i}}~$6678$}
\newcommand{\Heoptb}{\hbox{{\rm He}\kern 0.1em{\sc i}}~$5876$}
\newcommand{\Heoptc}{\hbox{{\rm He}\kern 0.1em{\sc i}}~$4471$}
\newcommand{\Brgam}{\hbox{{\rm Br}\kern 0.1em$\gamma$}}
\newcommand{\Brten}{\hbox{{\rm Br}\kern 0.1em$10$}}
\newcommand{\Breleven}{\hbox{{\rm Br}\kern 0.1em$11$}}
\newcommand{\HeIh}{\hbox{{\rm He}\kern 0.1em{\sc i}}~$1.7$~{\micron}}
\newcommand{\HeIk}{\hbox{{\rm He}\kern 0.1em{\sc i}}~$2.06$~{\micron}}
\newcommand{\squishlist}{
   \begin{list}{$\bullet$}
    { \setlength{\itemsep}{0pt}      \setlength{\parsep}{1pt}
      \setlength{\topsep}{3pt}       \setlength{\partopsep}{0pt}
      \setlength{\leftmargin}{1.5em} \setlength{\labelwidth}{1em}
      \setlength{\labelsep}{0.5em} } }
\newcommand{\squishend}{
    \end{list}  } 
\usepackage{url}

\slugcomment{ApJ in press; accepted 17 May 2009}
\shorttitle{[O IV] 26 micron as an intrinsic AGN luminosity measure}
\shortauthors{Rigby \etal}

\begin{document}
\title{Calibration of [O IV] 26 \micron\ as a Measure of Intrinsic AGN Luminosity}
\author{J.~R.~Rigby\altaffilmark{1,2}, A.~M.~Diamond-Stanic\altaffilmark{3}, \& G.~Aniano\altaffilmark{4}}
\altaffiltext{1}{Observatories, Carnegie Institution of Washington}
\altaffiltext{2}{Spitzer Fellow}
\altaffiltext{3}{Steward Observatory, University of Arizona}
\altaffiltext{4}{Department of Astrophysical Sciences, Princeton University}
\email{jrigby@ociw.edu}

\begin{abstract}
We compare [O IV] 25.89~\micron\ emission line luminosities with 
very hard (10--200~keV) X-rays from Swift, Integral, and BeppoSAX 
for a complete sample of 89 Seyferts from the Revised Shapley--Ames sample.  
Using Seyfert 1s, we calibrate [O IV] as a 
measure of AGN intrinsic luminosity, for particular use in high--obscuration environments.
With this calibration, we measure the average decrement in 14--195 keV X-ray to [O IV] 
luminosity ratio for Seyfert 2s compared to type 1s.
We find a decrement of $3.1 \pm 0.8$ for Seyfert 2s, 
    and a decrement of $5.0 \pm 2.7$ for known Compton-thick Seyfert 2s.  
These decrements imply column densities of approximately 
$\log N_H = 24.6$~\cmsq\ and $24.7$~\cmsq, respectively.  
Thus, we infer that the average Seyfert 2 is more highly obscured and intrinsically
more luminous than would be inferred even from the very hard X-rays.
We demonstrate two applications of the hard X-ray to [O IV] ratio. 
We measure a column density for the extremely obscured NGC 1068
of $\log N_H = 25.3$--25.4~\cmsq.  
Finally, by comparing [O IV] luminosities to total infrared luminosities for twelve bright
ultraluminous infrared galaxies, we find that four have substantial AGN contributions.
\end{abstract}


\section{Introduction} \label{sec:intro}
The optical line [O III] 5007~\AA\ is widely used to measure the intrinsic luminosity of active
galactic nuclei (AGN) \citep{heckman05}.  
However, there are cases, such as Seyfert 2s, ultra-luminous infrared galaxies, 
and Compton--thick AGN, where the extinction along our line of sight to the narrow line regions may be high,
and a more extinction--robust diagnostic may be needed.
\citet{melendez} recently demonstrated a correlation in AGN 
between the [O IV] 25.89~\micron\ emission line luminosity 
and the very hard X-ray luminosity (14--195 keV) as measured by the BAT instrument on Swift.
The sample of 40 AGN in Melendez \etal, selected by hard X-ray--flux,
was sufficient to demonstrate a correlation, but insufficient to calibrate the relationship.
In a companion paper to this one \citep{aleks}, we find that Seyfert 1s and Seyfert 2s from
a complete sample  
have statistically indistinguishable [O IV] luminosity distributions,
which further supports [O IV] as an AGN luminosity diagnostic.  
Here we compare the [O IV] and hard ($E>10$ keV) X-ray luminosities of a complete sample of 89 Seyferts,
bolometrically correct the hard X-rays to infer intrinsic AGN luminosities, 
and calibrate [O IV] as a measure of intrinsic AGN luminosity.


One potential problem with [O IV]~26~\micron\ is that its critical density of
$\log n_{crit} = 4.06$~\cmsq\ is lower than, for example, that of  
[Ne V]~14.32~\micron\   ($\log n_{crit} = 4.70$~\cmsq), 
[Ne V]~24.31~\micron\   ($\log n_{crit} = 4.44$~\cmsq), 
or [O III] 5007~\AA\ ($\log n_{crit} = 5.8$~\cmsq)\footnote{[O IV] and [Ne V] critical densities
were calculated at $T=10^4$~K by Cloudy, v08.00, last described by \citep{ferland}.   
The [O III] critical density at $T=10^4$~K is from \citet{osterbrock_book}.}.
Aniano \etal\ (in prep.) find, using the same Seyfert sample as this paper, 
that the [Ne V] 14 / [Ne V] 24~\micron\ line ratio is almost always $<$2.0, which
for temperatures of $\le 3\times 10^4$~K implies densities below $10^4$~cm$^{-3}$.  
Thus, [Ne V]/[Ne V]  indicates that the densities are below the  [O IV]~26~\micron\  critical density, 
though this measurement applies only to the [Ne V]--emitting portion of the narrow--line region.  

Another potential problem, which affects any intrinsic luminosity diagnostic based on emission lines 
from the narrow--line region (NLR), is that one must assume that the NLR enjoys an unobscured 
line of sight down to the central engine.  If this assumption is violated, then the NLR will be ionized by
less than the true continuum strength, 
and consequently NLR diagnostics will under-estimate the intrinsic AGN luminosity. 

For now, we develop [O IV] as a luminosity indicator, with the caveats
that it may cease to work when densities exceed critical, and that like [O III] 5007\AA\ it assumes
the narrow line region has not been shielded from the full intensity of the central engine.
In \S\ref{sec:ulirgs}, we examine [O IV] luminosities 
in ultra-luminous infrared galaxies (ULIRGs), which as deeply--obscured galaxies may 
be worst--case environments to test the [O IV] diagnostic.

\section{The Sample and the Data}  \label{sec:data}

Calibration of [O IV] against hard X-ray luminosity is
best done within a complete sample to avoid flux biases.  
By contrast, the members of the Melendez \etal\ sample used to establish the correlation between 
[O IV] and hard X-ray were selected as bright Swift detections, and therefore may well be biased 
toward X-ray--brightness.
Our sample, which is limited by host galaxy magnitude,
is the set of 89 Seyferts from \citet{MR95} and \citet{ho97} with B$_T \le 13$.
This Revised Shapley--Ames (RSA, \citealt{SA}) Seyfert sample is described in more detail by \citet{aleks}.  
We take distances and classifications from \citet{aleks}; the median distance is 22~Mpc.  
Optical spectral classifications of Seyfert 1--1.5 we group as ``Seyfert 1'', and
we group Seyfert 1.8--2 classifications as ``Seyfert 2''.
From column densities reported in the literature from X-ray measurements, we classify the AGN
into reported Compton--thin, reported Compton--thick, and 
Seyferts whose column densities are not reported in the literature.


Spectra were obtained by the Infrared Spectrograph (IRS) onboard Spitzer.  We use [O IV] fluxes from the LL1 order,
as measured and tabulated by \citet{aleks}.  
These fluxes have been corrected for contamination from the [Fe II] 25.99~\micron\ emission line, 
as described by \citet{aleks}.  
For the few AGN with undetected [O IV], we use the $3\sigma$ upper limit.
For the very bright sources NGC 1068 and Circinus,  we take ISO/SWS fluxes from \citet{sturm},
and for NGC 4945 from \citet{spoon}.

We take hard X-ray fluxes from published catalogs:  
\begin{itemize}
\item Swift BAT (14-195 keV) fluxes from the 22 month survey \citep{tueller09};  
\item BeppoSAX PDS (20-100 keV) fluxes from \citet{dadina}; and 
\item Integral IBIS (17-60 keV) fluxes from \citet{krivonos}.
\end{itemize}
By using  all three hard X-ray satellites,
we obtain meaningful upper limits (since the Swift survey was all-sky),
probe to fluxes fainter than the Swift all-sky detection limit,
and gauge the importance of X-ray variability.
Since Integral and BeppoSAX observations were targeted, 
the depth of coverage is highly non--uniform across the sky, 
so we do not estimate upper limits on non-detections.
Since the Swift/BAT survey by design has fairly uniform all-sky coverage, 
for non-detections we plot upper limits of $3.1 \times 10^{-11}$ \cgsflux, 
which is the $4.8 \sigma$ depth the 22 month Swift survey reached for $90\%$ of the sky \citep{tueller09}.  

We scale BeppoSAX fluxes to equivalent Swift fluxes using the bolometric corrections 
calculated in \S\ref{sec:bolocor} for the \citet{marconi04} template; 
for AGN detected by both instruments the scaled fluxes agree within errors.  
We scale Integral fluxes by an empirical factor of $\times 1.61$, 
rather than the factor of $\times 1.99$ derived from the Marconi template.  
This empirical factor is the best--fit linear relation for RSA Seyferts
detected by both Integral and Swift.\footnote{After dropping the two weakest detections.}  
The offset from the predicted relation is not surprising,
since the Integral fluxes assume the spectral shape
of the Crab nebula, which is softer than AGN over this spectral range
(the Crab nebula has $\Gamma = 2.14$ over the 1--700 keV range according to \citealt{kuiper-crab}.)

\section{The Bolometric Correction}  \label{sec:bolocor}
One wants to consider the total amount of ``direct'' or ``intrinsic'' AGN emission, 
which is radiated primarily in the X-ray, UV, and optical.  
This is the luminosity inferred to have been emitted into $4\pi$ steradians before any obscuration.  
One should separately consider emission that has been absorbed and re-radiated into the infrared, since
this will depend on the geometry of the absorbing material.  
A so-called bolometric correction, based on model spectral energy distributions, is used to translate
a luminosity measured at a particular band (for example, hard X-rays) into the 
intrinsic luminosity of the AGN.   (This is termed a ``bolometric correction'' in the literature, even 
when, as in this case, the reprocessed infrared emission is deliberately ignored.)
\citet{marconi04}  constructed a template spectrum using broken power laws in the optical--UV, 
a Rayleigh--Jeans blackbody tail in the near-IR, and a power-law spectrum with 
$\Gamma = 1.9$ plus a reflection component in the X-ray.  
The X-ray luminosity was scaled to match the 
luminosity dependence of $\alpha_{OX}$ measured by \citet{VBS}.  
The resulting templates have lower X-ray--to--intrinsic luminosity ratios than \citet{elvis94},
as expected since the \citet{elvis94} sample was chosen to be X-ray bright.  
In correcting for that bias, \citet{elvis02} shifted the X-ray portion of their template down 0.135 dex;
this accounts for all but 0.05 dex of the offset between the \citet{elvis94} and \citet{marconi04} templates.

From these templates, Marconi \etal\ calculate the conversion from L(2--10 keV) to L(intrinsic)
as a function of L(intrinsic), and plot the relation in their figure 3b.  
We simply scale this conversion to higher X-ray energies.  
Fitting a fifth-order polynomial to the Marconi et al. template and 
integrating over the relevant X-ray bands, the scaling relations are:
\begin{equation} f(14-195~keV) / f(2-10~keV) = 2.67 \end{equation}  
\begin{equation} f(20-100~keV) / f(2-10~keV) = 1.74 \end{equation}  
\begin{equation} f(17-60~keV)  / f(2-10~keV) = 1.34 \end{equation} 
for the Swift BAT, BeppoSAX PDS, and Integral IBIS bands, respectively.
We then fit the dependence of intrinsic luminosity L$_I$ on 2--10 keV luminosity L$_{2-10}$  
in the Marconi models as:
\begin{eqnarray} 
\log (L_I / L_{\odot}) &=& 0.03776~[\log (L_{2-10}/L_{\odot})]^2  +\\ 
                       & & 0.5340 \log (L_{2-10}/L_{\odot}) +  2.276 \nonumber
\end{eqnarray}
This equation, along with the scaling relations in eqn. 1--3, can be used to 
infer an intrinsic AGN luminosity from a measured high-energy X-ray luminosity.

\begin{figure}
\figurenum{1}
\includegraphics[width=3.5in,angle=270]{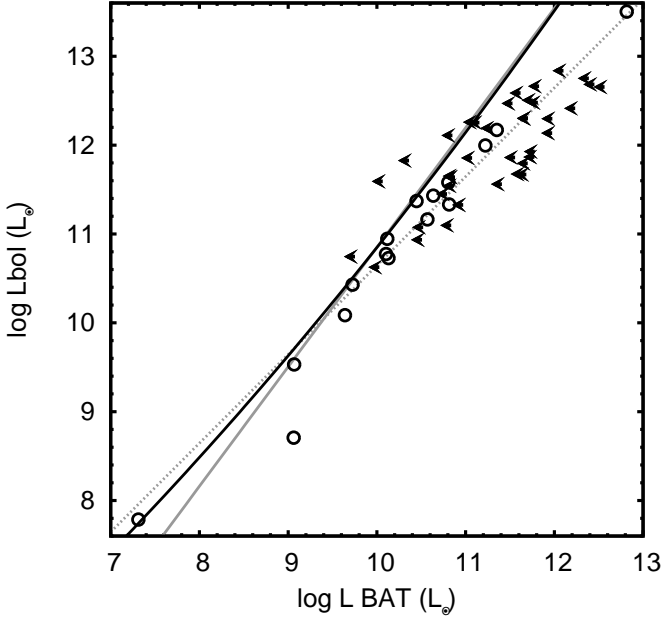}
\figcaption{Comparison of bolometric luminosity (inferred from FUSE far-UV) 
and Swift BAT hard X-ray luminosity for the 16 AGN in \citet{VF07} detected by BAT.  
Two fits are plotted:  a proportional fit to the detections  \textit{(dashed line)} 
of $L_I$/\Lsun $= 4.42$ L$_{BAT}$;
and a non-linear fit that includes the BAT upper limits using the Buckley-James regression
method \textit{(grey solid line)} 
of $\log(L_I / L_{\odot}) = 1.345 \log (L_{BAT} / L_{\odot}) - 2.60$. 
The non-linear fit agrees extremely well with the relation from \citet{marconi04} 
\textit{(black solid line)}, where a bolometric correction
has been applied between BAT and 2--10~keV.
}
\label{fig:vasudevan}
\end{figure}

The \citet{marconi04} template is well--suited to this application because it accounts for 
the known luminosity dependence of $\alpha_{OX}$ \citep{VBS}, and because it extends in
energy up to 1 MeV, unlike for example the \citet{elvis94} template which stops at 40 keV.  
We have already shown that the \citet{marconi04} template agrees  well with the
revised Elvis \etal\ template (2002), and will thus generate similar bolometric corrections.  
We now check the bolometric correction against \citet{VF07}.
We consider the 54 AGN from \citet{VF07} that have intrinsic AGN luminosities 
estimated from FUSE.  Sixteen are detected in the \citet{tueller09} Swift catalog. 
In Figure~\ref{fig:vasudevan} we plot the AGN luminosity from \citet{VF07} 
against the Swift BAT luminosity.  We plot a simple linear fit considering only detections,
as well as a non-linear fit that includes BAT upper limits using the Buckley-James regression
method \citep{isobe}.     Figure~\ref{fig:vasudevan} shows that the non-linear fit agrees 
extremely well with the Marconi relation.  

\section{Comparing [O IV] and hard X-ray luminosities} \label{sec:OIVX}

Figure~\ref{fig:swift-OIV} plots Swift--equivalent luminosity against [O IV] luminosity.
Each detection is plotted separately,
with symbols to indicate the satellite (squares for Swift, circles for BeppoSAX, triangles for Integral).
(For example, an AGN detected by all three satellites is represented by a cluster of three points at the same 
L[O IV] value.)  
The AGN are split into four panels using classifications described in \S\ref{sec:data}:
 Seyfert 1; Compton-thin Seyfert 2;
Seyfert 2 with unknown column density; and Compton--thick Seyfert 2.

Next, we use the bolometric corrections of \citet{marconi04} to convert the 
Swift--equivalent X-ray luminosities into intrinsic AGN luminosity, 
and plot against L[O IV] in Figure~\ref{fig:Ltot-OIV}.  
Symbols indicate column density and spectral classification.
We confirm the result of Melendez \etal\ (2008) that Seyfert 2s have 
systematically lower X-ray/[OIV] luminosity ratios than Seyfert 1s.  
In addition, with our larger sample of 18 Compton--thick AGN 
(compared to 4 in \citet{melendez}), of which 15 were detected in [O IV], 
and 6 were detected in hard X-ray, we show that Compton--thick AGN have still lower ratios.  

For each AGN class, we fit the ratio of intrinsic AGN luminosity inferred from the hard X-rays to
[O IV] luminosity using the Kaplan-Meier estimator of the survival function \citep{feigelson}, 
as implemented in the ASURV package \citep{asurv}.  This estimator incorporates both detections and upper limits
to provide a maximum likelihood reconstruction of the true distribution function, even in cases
where there are more upper limits than detections (as for the Compton-thick AGN.)

For AGN detected by multiple X-ray instruments, we use the average scaled X-ray flux.  
%
For Seyfert 1s, the best-fit ratio is         $2550 \pm 740$ (with a scatter of 0.4 dex);
for     Seyfert 2s              it is          $810 \pm 170$;
for known Compton--thin Seyfert 2s it is      $1320 \pm 300$;
and for known Compton-thick Seyfert 2s it is   $510 \pm 270$.\footnote{We exclude NGC 1068 
from the fit because its ratio is so low, and consider it separately in \S\ref{sec:n1068}.}
If we exclude the anomalously X-ray--bright NGC 4945 from the fit, the best-fit ratio for the 
Compton-thick Seyfert 2s falls to $240 \pm 70$.\footnote{Excluding NGC 1068 as well.}
Thus, Seyfert 2s have an average hard X-ray to [O IV] ratio that is $3.1 \pm 0.8$ times lower than Seyfert 1s.  
Known Compton-thin Seyfert 2s show a decrement of $1.9 \pm 0.5$, 
and known Compton-thick Seyfert 2s show a decrement of $5.0 \pm 2.7$ (excluding NGC 1068.)  

The two--sample ASURV statistical tests find a $<0.06\%$ probability that the X/[O IV] ratios 
for Seyfert 1s and Seyfert 2s are drawn from the same population.  
It is similarly improbable ($<0.09\%$)
that Seyfert 1s and Compton-thick Seyfert 2s are drawn from the same population.  
However, the offset between S1s and Compton-thin Seyfert 2s is of lower statistical significance 
($9\%$ probability of being drawn from the same population).

\section{Estimating column densities}\label{sec:column}

\begin{figure}
\figurenum{2}
\includegraphics[width=2.1in,angle=270]{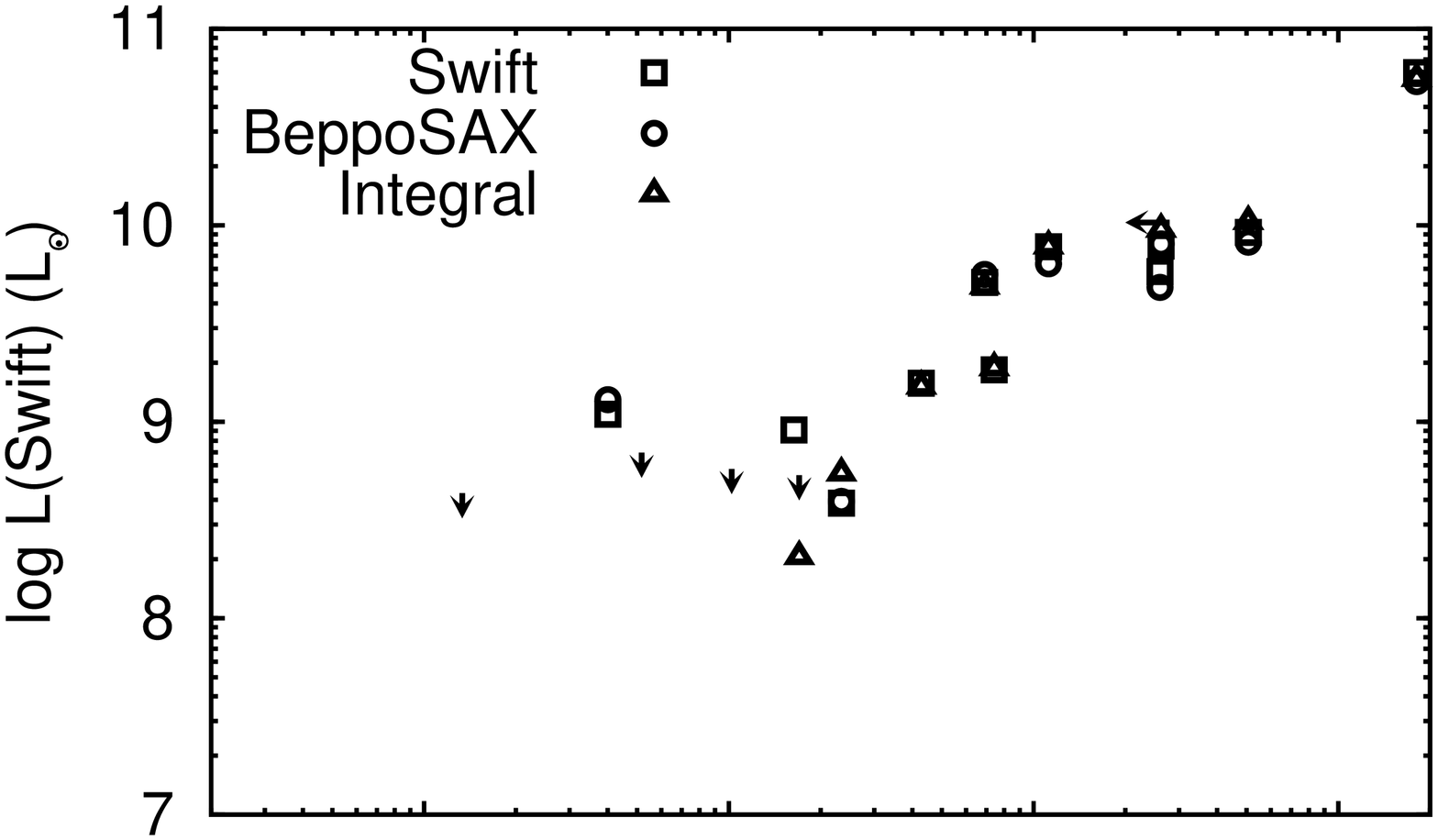}  

\vspace{-0.3in}

\includegraphics[width=2.1in,angle=270]{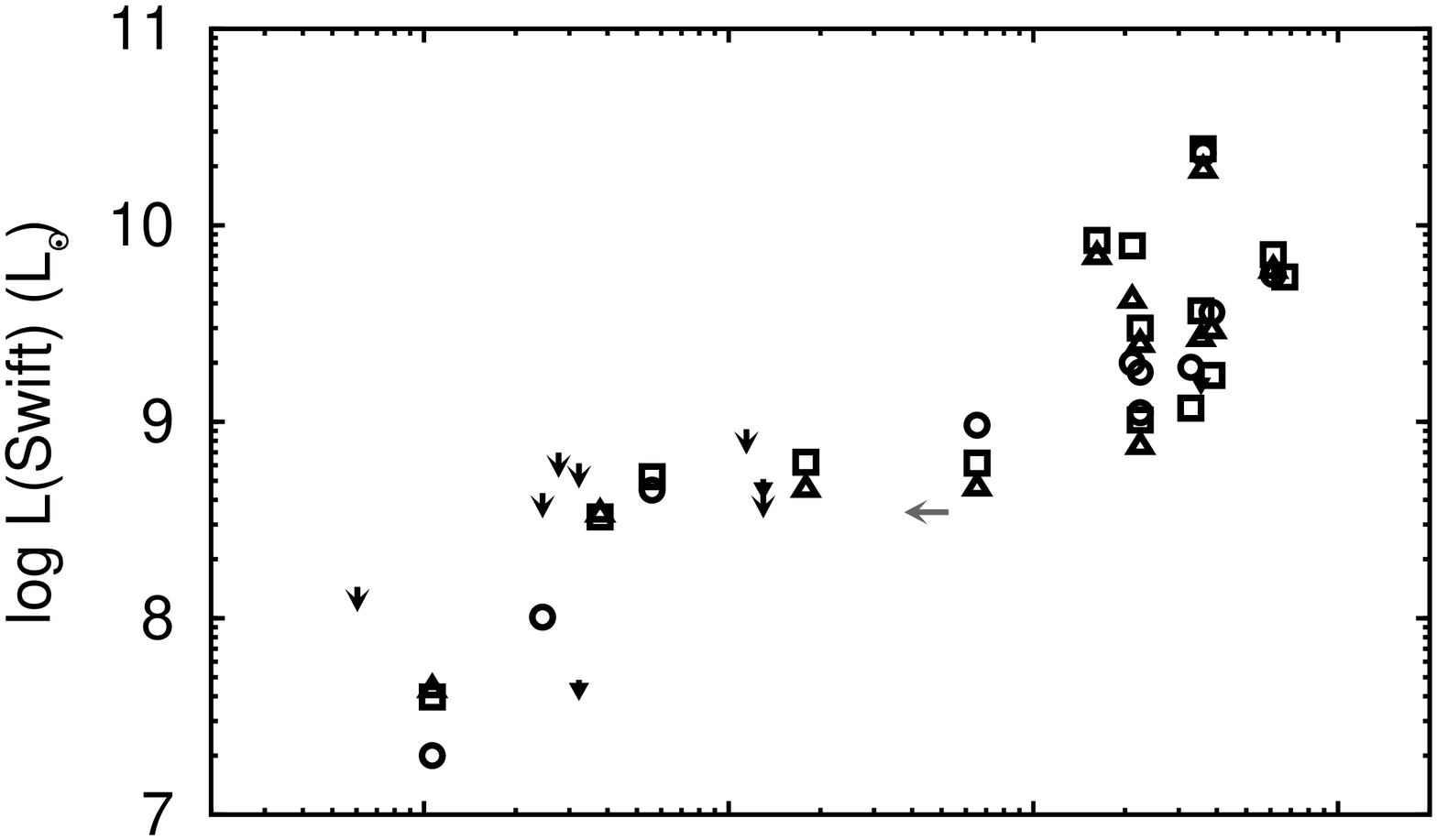}

\vspace{-0.3in}

\includegraphics[width=2.1in,angle=270]{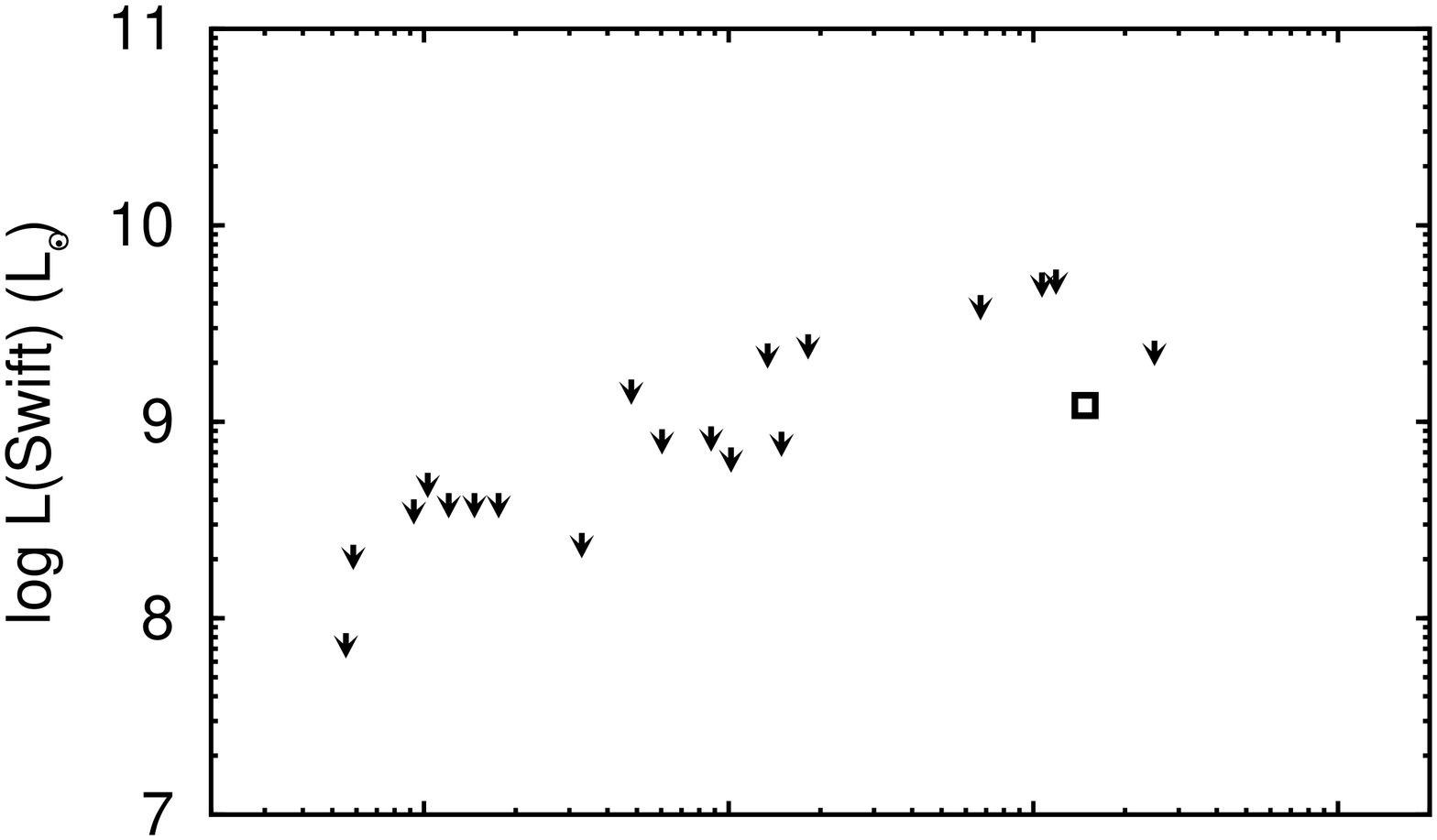}

\vspace{-0.3in}

\includegraphics[width=2.566in,angle=270]{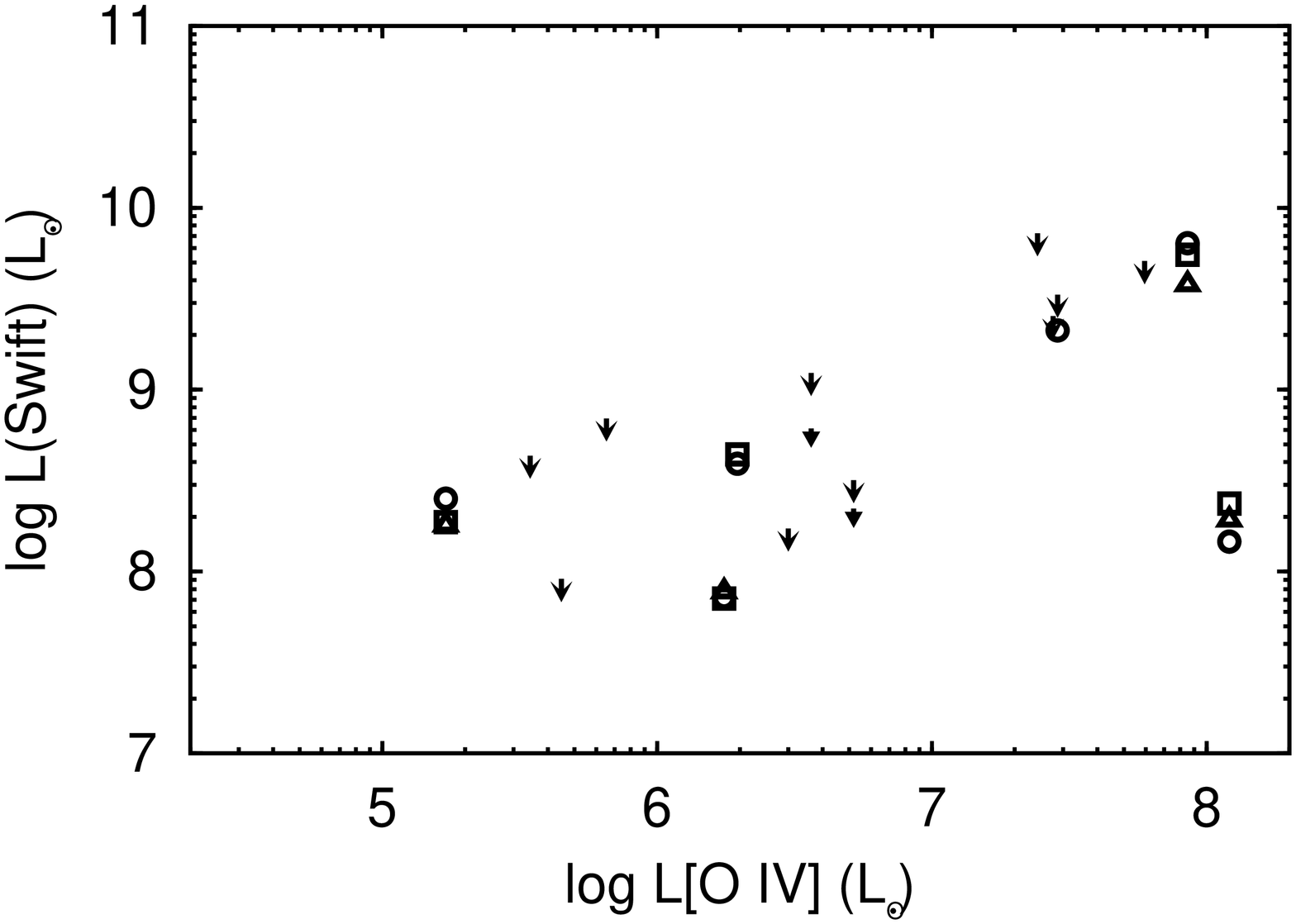}
\newpage
\figcaption{Hard X-ray luminosities as a function of [O IV] line luminosity.  
Seyfert 1 AGN are plotted in the top panel; 
known Compton-thin Seyfert 2 in the second panel;
Seyfert 2 of unknown column density in the third panel;
and Compton--thick Seyfert 2 in the bottom panel.
Swift BAT detections are squares (upper limits are black partially--filled arrows);
BeppoSAX PDS detections are circles (upper limits are black filled arrows);
and Integral detections are triangles (upper limits are grey arrows.)
BeppoSAX and Integral fluxes have been scaled to equivalent fluxes in the Swift BAT bandpass.
}
\label{fig:swift-OIV}
\end{figure}

\begin{figure}  
\figurenum{3}
\includegraphics[width=7in,angle=270]{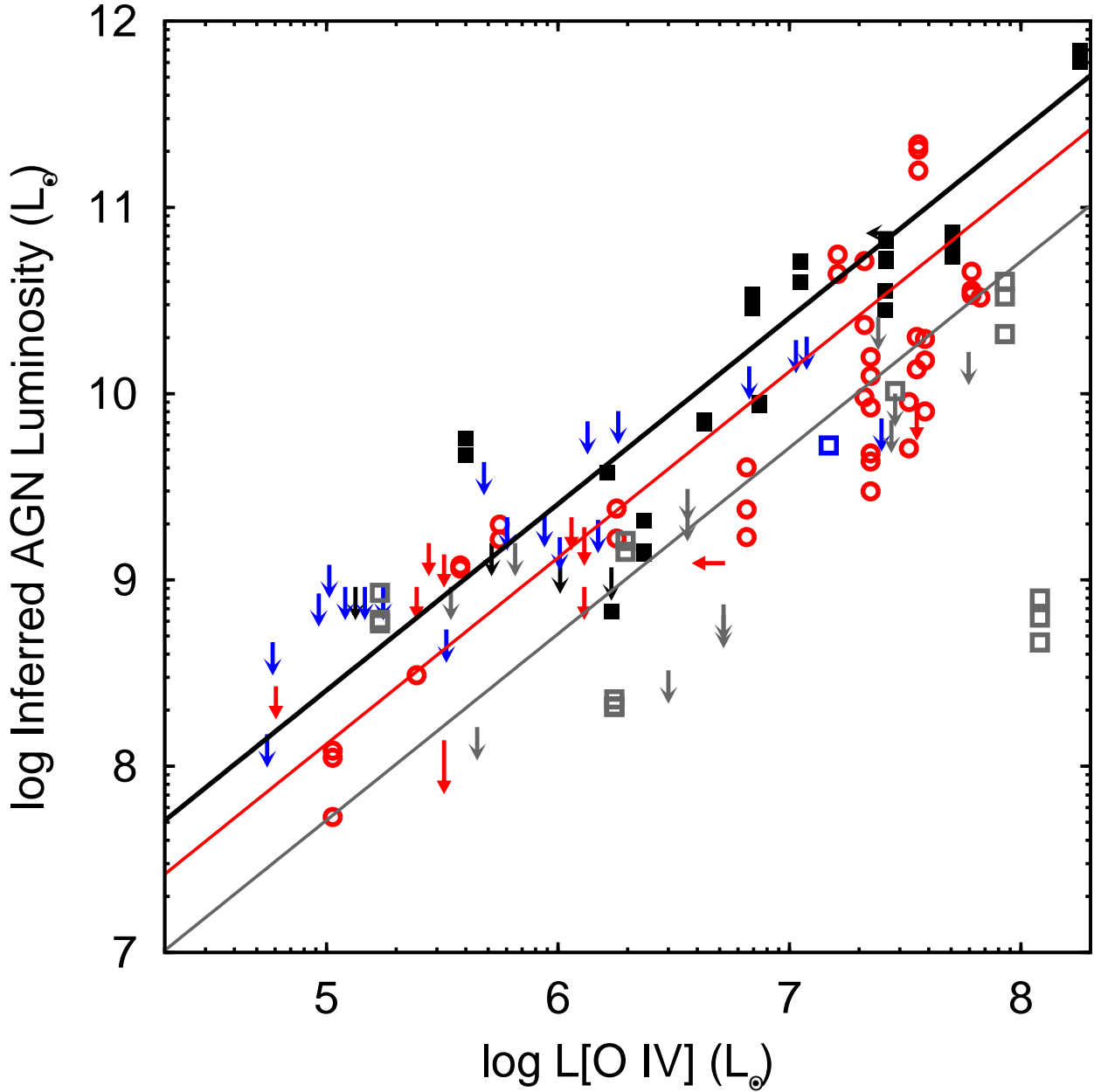}     
\figcaption{Intrinsic AGN luminosity, as inferred from the hard X-rays, versus [O IV] luminosity.
Seyfert 1--1.5 AGN are plotted in black as squares (detections) and arrows (non-detections);
Compton-thin Seyfert 1.8--2 AGN are plotted as red open circles and arrows; 
Seyfert 1.8--2 of unknown column are plotted as blue arrows or squares;
and Compton-thick AGN are plotted as light grey open squares and  arrows.
The best fit X-ray--inferred intrinsic to [O IV] luminosity ratio is plotted for each AGN class, 
in the same color scheme; the fits are given in \S\ref{sec:OIVX}.} 
\label{fig:Ltot-OIV}
\end{figure}

We now interpret the difference in hard X-ray/[O IV] ratio between type 1 and type 2 Seyferts.
One explanation would be that the [O IV] luminosities for the type 1 Seyferts are biased high. 
In a companion paper \citep{aleks}, we find no significant differences between the [O IV] 
luminosity distributions of type 1 and type 2 Seyferts; the same conclusion can be reached 
by inspecting Figures \ref{fig:swift-OIV} and \ref{fig:Ltot-OIV}.  
Thus, there is no evidence for a bias toward high [O IV] luminosities.

Therefore, we use L[O IV] as a measure of intrinsic AGN luminosity, and 
interpret the offset in hard X-ray/[O IV] ratios as caused by substantial absorption
in the E$>$10~keV spectra of  Seyfert 2s, as suggested by \citet{melendez}.
At such high energies, the cross-section for Compton scattering is larger than that
of photoelectric absorption.  Consequently, one expects a typical high--energy
photon to experience multiple Compton scatterings, losing energy each time to electron recoil,
until the photon either escapes, or loses sufficient energy and traverses a sufficient path
that it is absorbed photoelectrically.  This makes the problem relativistic,
optically--thick, and geometry-dependent, and as such, Monte Carlo simulations are required to 
interpret the flux decrement.  Fortunately, \citet{matt99} simulated exactly this situation.  
We numerically integrate their spectra over the Swift/BAT, BeppoSAX/PDS, and 
Integral/IBIS energy bands and compare to the injected spectrum; 
the resulting flux decrements are listed in Table~\ref{tab:matt99}.

In the \citet{matt99} model, suppressing a  BAT flux 
by a factor of $3.1 \pm 0.8$ (as observed for all Seyfert 2s in our sample) 
requires a column density of 
         $\log N(H) = 24.6^{+0.1}_{-0.2}$~\cmsq.  
Suppression by a factor of $5.0 \pm 2.7$ (as observed for known Compton-thick Seyfert 2s\footnote{Excluding NGC 1068.})
requires $\log N(H) = 24.7^{+0.2}_{-0.3}$~\cmsq.
These results are consistent with the column density of $\log N(H) = 24.5\pm0.1$ inferred by 
\citet{melendez} for their sample of 17 Seyfert 2s.
Thus, the [O IV] to hard X-ray ratio confirms that, on average, Seyfert 2s are more obscured 
in the hard X-rays than Seyfert 1s, and predicts the obscuring column densities expected for 
Compton-thick AGN.  

Surprisingly, the average ratio for the known Compton-thin Seyfert 2s also suggests high columns.
Suppression by a factor of $1.9 \pm 0.5$ can be explained by $\log N(H) = 24.3^{+0.1}_{-0.3}$~\cmsq.  
By contrast, the median column density reported in the literature for these AGN, based on 2--10~keV 
measurements, is $\log N(H) = 23.0$~\cmsq\ \citep{aleks}.  As such, the hard X-ray/[O IV] ratios imply higher
columns (Compton-thick or nearly so) than inferred from the 2-10 keV spectra.  However,
this is not the only plausible interpretation.  
As discussed in \S\ref{sec:OIVX}, the offset in hard X-ray/[O IV] ratio between
Compton-thin Seyfert 2s and Seyfert 1s is of low statistical significance in our sample.  
As such, we cannot rule out Compton-thin Seyfert 2s having, on average, low columns.  
Table~\ref{tab:matt99} shows that column densities below $\log N(H) \sim 24$~\cmsq\ should produce only 
a small flux decrement.  Accordingly, for such low columns, soft X-ray spectra should return a more 
precise measurement of the column density.

Figure \ref{fig:Ltot-OIV} does show that a substantial minority of the Compton-thin Seyfert 2s 
have X/[O IV] ratios as low as known Compton-thick AGN.  In order of increasing X/[O IV] ratio, 
these are NGC 1365, NGC 7582, NGC 2992, NGC 7314, NGC 3081, and NGC 5728.\footnote{The remaining
Compton-thin Seyfert 2s have an average hard X-ray to [O IV] offset from the Seyfert 1s that is of
even lower statistical significance.}
There several possible explanations for these low ratios:  they may have nearly Compton-thick or Compton-thick
columns,  have column densities that vary dramatically with time, 
or have dramatic variations in hard X-ray flux.  
The literature reveals that all three explanations appear to be at work:
\begin{itemize}
\item NGC 1365 varies between Compton-thin and Compton-thick in a matter of days \citep{risaliti07};
\item NGC 7582 has experienced dramatic changes in its 2-10 keV flux consistent with a variable Compton-thick absorber \citep{picon};
\item NGC 2992 was measured to have a column of only $7\times 10^{21}$\cmsq\ \citep{colbert}, 
but its 2-10 keV flux has varied by a factor of 20 over twenty years \citep{gilli00}, 
and its BAT flux has varied by a factor of 6 \citep{beckmann};
\item NGC 7314 was measured to have a low column density \citep{risaliti02};
\item NGC 3081 was measured to have a column density of $6\times10^{23}$~\cmsq\ \citep{maiolino98}, 
      and varies in the BAT band by a factor of 7 \citep{beckmann};
\item NGC 5728 is borderline Compton thick with a column of $8\times 10^{23}$~\cmsq\ \citep{zhang06}.
\end{itemize}
Thus, five of these six AGN have a plausible explanation for their low X/[O IV] ratios; in four
cases the explanation appears to be that the column is almost Compton-thick, or transitions between 
Compton thick and thin.  This comparison also suggests that an inherent drawback of calibrating [O IV] 
against the hard X-rays is that the hard X-ray flux may vary with time.
Still, the systematically low X/[O IV] ratios observed for all Seyfert 2s in this sample strongly suggest
that the typical Seyfert 2 has a significantly attenuated hard X-ray flux, and is 
considerably more powerful than would be inferred from the hard X-rays.

\section{Application: NGC 1068}\label{sec:n1068}
The famous Compton--thick NGC 1068 is the most highly--obscured AGN in our sample, based on its
hard X-ray to [O IV] luminosity ratio.  
Its [O IV] luminosity\footnote{Assuming a distance of 14.4 Mpc.}  
of $1.2\times 10^{8}$~\Lsun\ predicts an intrinsic AGN luminosity of  
($3.1 \pm 0.9$) $\times 10^{11}$~\Lsun\ using the Seyfert 1 calibration.           
The [O IV] flux is taken from high resolution ISO spectra \citep{sturm}, so 
[Fe II] contamination is not an issue; in any case the [Fe II] line strength 
is only $4\%$ that of [O IV].
The hard X-ray to [O IV] ratio is  $490\times$ lower than the Seyfert 1 best fit,
implying a column density of $\log N_H = 25.3$ to 25.4~\cmsq\ in the
\citet{matt99} models.  (The column should be still higher if scattered light contributes substantially 
to the observed X-ray flux.)  This inferred column is consistent with the lower limits inferred through a similar
comparison to [O III] 5007\AA\ \citep{matt97, matt2000}.
In a future paper, we plan to compare this inferred intrinsic luminosity in detail with
the reprocessed mid-infrared SED.

 

\section{Application: ultra-luminous infrared galaxies}\label{sec:ulirgs}
As an application of the [O IV] calibration, we consider the [O IV] luminosities for 
twelve nearby ultra-luminous infrared galaxies (ULIRGs), 
using data from \citet{armus04} and \citet{armusBGS}, which 
comprises the ten brightest ULIRGs plus two others.  
Only five ULIRGs have detected [O IV]:  
IRAS 05189-2524, IRAS 13428+5608 (Mrk 273), IRAS 09320+6134 (UGC 5101), 
Mrk 1014, and Mrk 463e.    
Using the relation established above for Seyfert 1s, 
we infer intrinsic AGN luminosities from  L[O IV]
and compare to the total infrared (8--1000~\micron) luminosities.
The inferred L(AGN)/L(IR) percentages are $49\%$, $97\%$, $22\%$, 
$170\%$, and $520\%$, respectively.  
(A $30\%$ errorbar should be applied for the uncertainty in the Seyfert 1 calibration.) 
For ULIRGs with undetected [O IV] we infer percentages below $8\%$.

Based on excitation diagrams using [Ne V]/[Ne II], [O IV]/[Ne II], and the equivalent width of the 
6.2\micron\ PAH feature, \citet{armusBGS} picked out 
three of these (Mrk 463e, Mrk 1014, and IRAS 05189-2524) as being likely
AGN--dominated.  In addition, they estimate Mrk 273 has an  AGN contribution of 50--70$\%$.  
UGC 5101, which [O IV] indicates has a modest AGN contribution, 
does not stand out as AGN-dominated in the \citet{armusBGS} excitation diagrams; however, 
it does have detected Fe K $\alpha$ with an equivalent width of 400 eV \citep{imanishi03}, 
which signals that an obscured AGN is present and may possibly be bolometrically important.
Thus, the L[O IV]/L(IR) ratio gives results consistent with the excitation diagrams as to which
ULIRGs are AGN--dominated.

Does the normalization of L[O IV] to L(AGN), calibrated from Seyfert 1s, appear roughly correct for ULIRGs? 
The preliminary answer appears to be yes, since the inferred AGN fractions are 
of order unity for four of the five the [O IV]--detected ULIRGs, 
consistent with the results of the excitation diagrams.  
\citet{armusBGS} did note that the AGN fractions implied by [Ne V]/[Ne II] or [O IV]/[Ne II]
line ratios were significantly lower than implied by the mid-IR slope or the 6.2 \micron\
PAH feature equivalent width; we do not see this effect for L[O IV]/L(IR).
The one clear outlier in  L[O IV]/L(IR) is Mrk 463e, for which we infer 
an L(AGN)/L(IR) ratio significantly above unity.    
Such a result is physically plausible if most of the AGN energy is not reprocessed into the infrared, 
but escapes via the X-ray/UV/optical.  
Indeed, the infrared luminosity for Mrk 463e is only $5\times10^{11}$~\Lsun, below the ULIRG cutoff, 
but its  total luminosity is considerably higher, well above the ULIRG threshold \citep{armus04}.  
In addition, Mrk 463e has remarkably weak PAH features, which argues that the starburst
contribution is small.
Thus, given that Mrk 463e is apparently an AGN--dominated ULIRG with a low ratio of reprocessed to
intrinsic light, the high measured ratio of L[O IV]/L(IR) seems reasonable and physical.

It remains unclear how to interpret the [O IV] non-detected ULIRGs, 
especially those like Mrk 231, which is a broad-line AGN. 
One explanation is that these galaxies are indeed dominated bolometrically by star formation, not AGN.  
An alternate explanation would be that they have bolometrically--important AGN, but that the [O IV] diagnostic
has failed, perhaps because densities exceed the [O IV] critical density, or because 
the narrow line regions have been shielded from the central engine.  
Thus, while the [O IV] diagnostic appears to work well for Seyferts in the RSA sample, 
and identifies what are thought to be the most AGN--dominated of bright ULIRG, 
more work is still required to understand its strengths and limitations of the diagnostic 
in extremely obscured environments.

\section{Conclusions}  \label{sec:conclude}

We present the first calibration of [O IV]~25.89~\micron\ as a measure of 
intrinsic AGN luminosity.  
The [O IV] line is bright, is detected in hundreds of AGN in the Spitzer archive, 
and will be detectable to Herschel and JWST.  

We find that Seyfert 2s show systematically lower ($E>10$~keV) ratios of X-ray to [O IV] luminosity than
Seyfert 1s, by a factor of $3.1\pm0.8$.  
For AGN previously identified as Compton-thick, the observed decrement is $5.0\pm2.7$.
We interpret this as caused by absorption that affects even the very hard X-rays.
Monte Carlo simulations \citep{matt99} associate these X-ray/[O IV] flux decrements 
with average column densities of   $\log N(H) = 24.6^{+0.1}_{-0.2}$~\cmsq\ and
$24.7^{+0.2}_{-0.3}$~\cmsq, respectively.
This, we find substantial obscuration in most Seyfert 2s.

We briefly explore use of [O IV] 26~\micron\ to estimate AGN power even in 
very obscured systems such as Compton-thick AGN and  ULIRGs.  For Compton--thick AGN
we infer sensible average column densities, and are able to infer a column density even for the
extremely obscured NGC 1068.
Application to ULIRGs is less straightfoward, 
since only five of the twelve bright ULIRGs we examine have detected [O IV].
For [O IV]--detected ULIRGs, 
the L(O IV)/L(IR) ratios imply significant AGN contribution.  
The question remains whether the non-detections have lower AGN fractions than the detections; it
is possible that the [O IV] diagnostic breaks down in the extreme environments of some ULIRGs.

We conclude by noting the recent interest in using very hard X-rays to find obscured AGN, 
due to their penetrating power.  Several sensitive, high--resolution $E>10$ keV 
wide-field or pencil-beam X-ray satellites are now in preparation or planning, 
with scientific goals of finding even the most obscured AGN in the nearby universe, 
and identifying the population of Compton--thick AGN presumed to contribute to the 
hardness of the X-ray background.  
Our results suggest that even the 20--200 keV band suffers considerable
flux attenuation in typical Seyfert 2s, especially those that are Compton-thick.  
Thus, our results predict that X-ray surveys will probe considerably smaller volumes for 
obscured AGN than for unobscured AGN, even when very hard energy bands are used. 
The [O IV] 26\micron\ line may play a critical role in measuring this absorption, and 
determining the true luminosities of even highly--obscured AGN.

\acknowledgments
Acknowledgments:  We thank G.~Matt for kindly providing Monte Carlo simulation results from
\citet{matt99} for additional column densities.  
JRR was supported by a Spitzer Space Telescope Postdoctoral Fellowship.

\clearpage

\begin{deluxetable}{llll}
\tablecolumns{4}
\tablewidth{0pc}
\tablenum{1}
\tablecaption{Ratio of emergent to input hard X-ray flux, from \citet{matt99} models.\label{tab:matt99}}
\tablehead{
\colhead{log N$_H$ (\cmsq)} &  \colhead{Swift/BAT} &  \colhead{SAX/PDS} & \colhead{Integral/IBIS} \\}
\startdata
23.0 &	 0.96   &	0.96  &	  0.96\\
23.5 &   0.88	&	0.89  &   0.87\\
23.8 &   0.78	&	0.80  &   0.77\\
24.0 &   0.69	&	0.71  &   0.68\\
24.1 &   0.64	&	0.66  &   0.63\\
24.2 &   0.58	&	0.61  &   0.57\\
24.3 &   0.51	&	0.54  &   0.50\\
24.4 &   0.44	&	0.48  &   0.44\\
24.5 &   0.37	&	0.41  &   0.38\\
24.6 &   0.30	&	0.34  &   0.31\\
24.7 &   0.23	&	0.27  &   0.25\\
24.8 &   0.17	&	0.20  &   0.19\\
24.9 &   0.11	&	0.13  &   0.13\\
25.0 &   0.063	&	0.077 &   0.074\\
25.1 &   0.048  &       0.058 &   0.056\\
25.2 &   0.028  &       0.034 &   0.033\\
25.3 &   0.0032 &       0.0039&   0.0039\\
25.4 &   0.0016 &       0.0020&   0.0020\\
25.5 &   0.00010&       0.00013&  0.00013\\
\enddata
\end{deluxetable}


\vspace{1in}


\end{document}